\documentclass[preprint,pra,aps,nofootinbib,longbibliography]{revtex4-1}

\usepackage{amsfonts}
\usepackage{graphicx}

\begin{document}

\title{A Holographic Principle for Non-Relativistic Quantum Mechanics}
\author{Russell B. Thompson}
\email{thompson@uwaterloo.ca}
\affiliation{Department of Physics \& Astronomy and Waterloo Institute for Nanotechnology, University of Waterloo, 200 University Avenue West, Waterloo, Ontario, Canada N2L 3G1}
\date{\today}

\begin{abstract}
The quantum-classical isomorphism for self-consistent field theory, which allows quantum particles in space-time to be represented as classical one-dimensional threads embedded in a five dimensional thermal-space-time, is summarized and used to explain a selection of quantum phenomena. Introduced by Feynman, and used for modern quantum simulations, the isomorphism, when phrased in a field-theoretic way, has been shown to be the same as quantum density functional theory, the theorems of which guarantee equivalent predictions with non-relativistic quantum mechanics. If the Feynman dimension is considered to be real, there is a duality between classical threads in five dimensions and quantum particles in four dimensions. Using the 5D picture, intuitive explanations are given for quantum phenomena including the uncertainty principle, tunnelling, geometric phase, and interference effects. Advantages of the 5D picture are presented, which include fewer postulates, no measurement problem, and the need for only classical concepts in the higher dimensional space. Limitations of the approach such as the interpretation of entanglement and spin are discussed.
\end{abstract}

\maketitle

\section{Introduction}
A holographic principle is an equivalence between the different mathematical descriptions of a system in a volume and on the surface of the volume. The term ``holographic'' is used because all the information about the system in the volume is contained on the surface, analogous to true optical holograms that encode information about a three-dimensional object on a flat surface. For a general holographic principle, the volume does not need to be three dimensional; a $d$ dimensional volume would contain the same information as the $d-1$ dimensional surface.  A famous example of a holographic principle is the anti-de Sitter/conformal field theory (AdS/CFT) correspondence \cite{Maldacena1998}. 

Richard Feynman introduced a mathematical trick by which quantum mechanical many-body problems can be solved using \emph{classical} statistical mechanics by treating the inverse of the thermal energy in the partition function as an imaginary time dimension \cite{Feynman1953a, Feynman1953b, Feynman1953c, Feynman1965}. This allows quantum simulations, such as path integral Monte Carlo \cite{Ceperley1995} or centroid \cite{Roy1999b, Zeng2014} and ring polymer \cite{Habershon2013} molecular dynamics, to be solved classically using an extra, fictitious, dimension. This has been called the quantum-classical isomorphism \cite{Chandler1981}.

In addition to simulations, it has been shown that polymer self-consistent field theory (SCFT) also obeys the quantum-classical isomorphism, and is, under the right conditions, equivalent to quantum density functional theory (DFT) \cite{Thompson2019, Thompson2022}. This allows one to treat the thermal dimension introduced by Feynman as more than just mathematics, and consider it a physical dimension giving a 5-dimensional thermal-space-time. A major motivation for doing this is that the theorems of DFT guarantee the equivalence between the DFT formalism and the predictions of quantum mechanics (QM) \cite{Hohenberg1964, Mermin1965, Runge1984}, allowing the explanation of all non-relativistic QM with fewer postulates \cite{Thompson2020, Thompson2022} that use only classical concepts, albeit in a higher dimensional classical space. 

SCFT represents quantum particles as classical one-dimensional threads embedded in the 5D thermal-space-time, and through the theorems of DFT, this model must make all the same predictions as QM \cite{Thompson2019, Thompson2020, Thompson2022}. It represents a holographic principle between QM in 4D (the surface) and polymer-like threads in 5D (the volume) which captures the wave-particle duality of QM. The 5D SCFT picture has fewer postulates than 4D wave function QM \cite{Thompson2020, Thompson2022}, and does not suffer the same pathologies as wave function QM. 

For example, since SCFT does not use wave functions, the concept of superposition is not necessary, and there is no wave function collapse. From the SCFT perspective, the measurement problem in 4D is an artifact of a projection onto the 4D surface which hides the physical 5D thread nature of quantum particles. There is no measurement problem or superposition in the 5D SCFT quantum particle picture any more than there is in SCFT for real polymers, which uses the same mathematics.

The origins of randomness in QM is also trivial in the 5D picture. SCFT is a statistical mechanics theory, and so in 5D, randomness has the same origins as in classical statistical mechanics. Specifically, there is ignorance about the conformations of the threads in the system. SCFT thus adheres to the ensemble interpretation of QM \cite{Ballentine1970}, in which QM only makes predictions about ensembles of systems.

SCFT mathematics predict, using a 5D classical framework without wave functions, the stability and shell structure of atoms \cite{Thompson2019, Thompson2020, LeMaitre2022}, spontaneous spherical symmetry breaking in atoms \cite{LeMaitre2023}, molecular bonding \cite{Sillaste2022}, and even the spontaneous emergence of classical electromagnetism \cite{Thompson2022}. In this paper, many of these features of SCFT, as presented at the 15th Biennial Quantum Structure Conference, are reviewed and other topics which arose in the questions and discussions following the presentation are explained, such as quantum statistics, the uncertainty principle, quantum kinetic energy, tunnelling, the double slit experiment, geometric phase including the Aharonov-Bohm effect, entanglement, exchange, and the Pauli exclusion principle. First principles derivations, numerical methods and more detailed discussions of SCFT applied to quantum systems can be found in references \cite{Thompson2019, Thompson2020, Sillaste2022, Thompson2022, LeMaitre2022, LeMaitre2023}. 

\section{Static Properties}
Consider a single quantum mechanical particle of mass $m$ at a temperature $T$. The quantum statistical mechanical partition function for this system is 
\begin{equation}
\mathcal{Q} = \sum_i e^{-\beta E_i}     \label{mcQ1}
\end{equation}
where $\beta = 1/k_BT$, $k_B$ is Boltzmann's constant and $E_i$ are the allowed energy states. Feynman showed that (\ref{mcQ1}) can be exactly rephrased as the path integral \cite{Feynman1953b}
\begin{equation}
\mathcal{Q} = \int d{\bf r}_0 \int_{tr} \exp\left\{ -\int_0^\beta \left[\frac{m}{2\hbar^2}\left(\frac{d{\bf r}}{ds}\right)^2+V({\bf r}(s))\right] ds \right\}\mathcal{D}{\bf r}   \label{mcQ2} 
\end{equation}
where $\hbar$ is Planck's reduced constant and $V$ is the potential. The integral $\int_{tr}$ is taken over all paths such that ${\bf r}(0) = {\bf r}(\beta) \equiv {\bf r}_0$, that is, closed ring paths. The partition function (\ref{mcQ2}) is identical with the classical partition function for a ring polymer in the Gaussian thread model of polymer self-consistent field theory (SCFT) \cite{Kim2012}. A ring polymer is represented in a coarse-grained way in SCFT as a mathematical contour ${\bf r}(s)$, embedded in three dimensional space $\mathbb{R}^3$. In (\ref{mcQ2}) however, $s$ refers to the inverse thermal energy $\beta$, and so is an independent dimension. Since SCFT is a \emph{classical} statistical mechanical theory, one can choose to picture quantum particles as classical one-dimensional threads in a space with an extra thermal dimension -- a five-dimensional thermal-space-time.

For many-body systems, SCFT uses a mean field approximation for a single particle subject to a field due to all other particles. For quantum systems, correlations and exchange effects can be included through the addition of an exchange-correlation term, just as in quantum DFT. In fact, as shown in appendix B of reference \cite{Thompson2019} and appendix C of reference \cite{LeMaitre2023}, polymer SCFT can be shown to be equivalent to Kohn-Sham DFT assuming perfect enforcement of the Pauli exclusion principle. Through the theorems of DFT \cite{Hohenberg1964, Mermin1965}, this means that all predictions of SCFT, again assuming the Pauli exclusion principle, must be consistent with those of non-relativistic quantum mechanics. In reality, approximations must always be made, and so discrepancies may arise, but in the non-interacting static case, one may expect agreement. 

For example, agreement should be perfect for particle-in-a-box situations. As expected, the SCFT solution of the hydrogen atom agrees exactly with the analytical quantum results \cite{Thompson2020, LeMaitre2022}. If correlations are ignored, the helium atom solved with SCFT agrees exactly with Hartree-Fock wave function theory \cite{LeMaitre2022}. Other quantum phenomena that can be explained intuitively in the static, non-interacting situation include the uncertainty principle, quantum kinetic energy, tunnelling and the measurement problem (collapse of the wave function). The equations of SCFT, as applied to quantum particles in the static case, are now reviewed in order to explain each of these phenomena. Complete derivational details can be found in reference \cite{Thompson2019}. 

\subsection{Summary of SCFT Equations}
\begin{figure}
\includegraphics[width=1.0\textwidth]{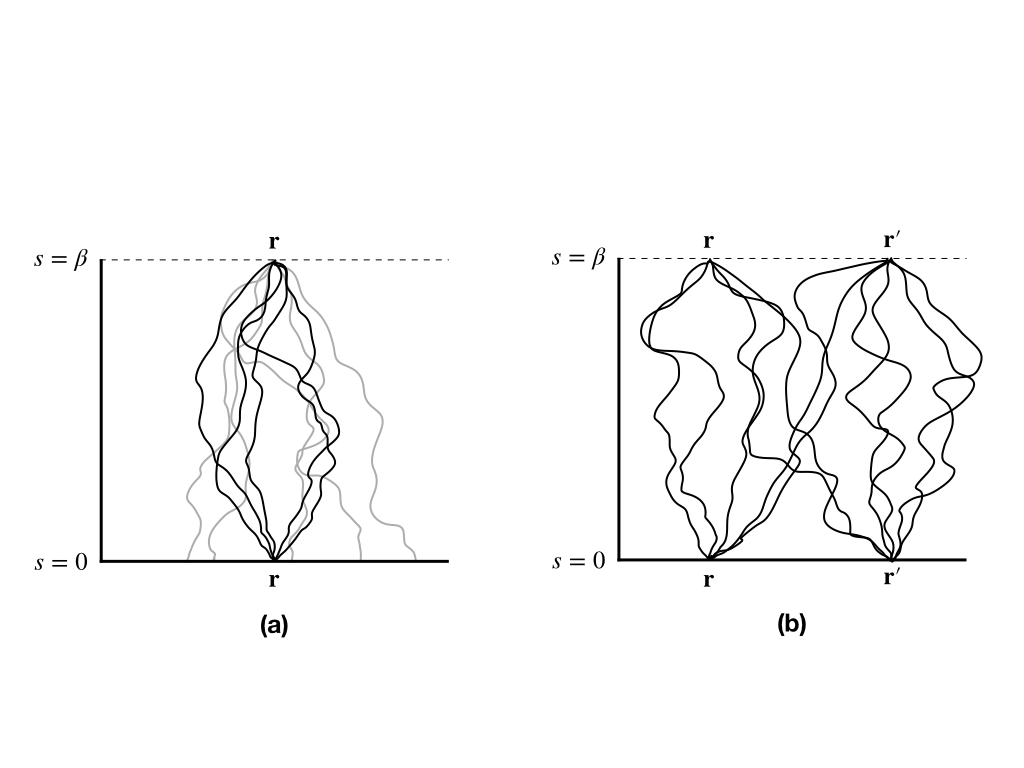}
\caption{Schematic of threads in thermal-space. $x$-axis is space and $y$-axis is inverse thermal energy $\beta$. (a) Contours that start from ${\bf r}$ and return to ${\bf r}$, forming rings, are shown in black; those that do not return to the starting position ${\bf r}$, forming open threads, are shown in grey. (b) Two starting positions ${\bf r}$ and ${\bf r}^\prime$ and some contours that terminate at ${\bf r}$ and ${\bf r}^\prime$. These include both rings and cross-paths that  start at ${\bf r}$ and end at ${\bf r}^\prime$ and vice versa.} \label{fig:threads}
\end{figure}

The main governing equation in SCFT is 
\begin{equation}
\frac{\partial q({\bf r}_0,{\bf r},s)}{\partial s} = \frac{\hbar^2}{2m} \nabla^2 q({\bf r}_0,{\bf r},s) - w({\bf r},\beta) q({\bf r}_0,{\bf r},s)  \label{diff1}
\end{equation}
subject to the initial condition
\begin{equation}
q({\bf r}_0,{\bf r},0) = \delta({\bf r}-{\bf r}_0)   .  \label{init1}
\end{equation}
This gives the unnormalized probability $q({\bf r}_0,{\bf r},s )$ that a particle at position ${\bf r}_0$ at high temperature ($s = 0$) will be found at position ${\bf r}$ at a temperature corresponding to $s=\beta$. Note that $q({\bf r}_0,{\bf r},s)$ is a completely real and positive definite quantity that depends on the field $w({\bf r},\beta)$ which quantifies the interactions due to all other quantum particles in the system and external potentials. For $N$ particles in the system, the single particle density at position ${\bf r}$ is then
\begin{equation}
n({\bf r},\beta) = \frac{N}{Q(\beta)} q({\bf r},{\bf r},\beta)  \label{dens1}  
\end{equation}
where the normalization factor 
\begin{equation}
Q(\beta) = \int q({\bf r},{\bf r},\beta) d{\bf r}   \label{Q1}
\end{equation}
is the partition function for a single particle subject to the field $w({\bf r},\beta)$. Equation (\ref{dens1}) shows that it is sufficient to consider ring polymers only, since only paths that start at ${\bf r}$ and return to ${\bf r}$ contribute to the quantum particle density at the position ${\bf r}$. That is, the density gives the probability of a particle at position ${\bf r}$ at high temperature ($s = 0$) returning to position ${\bf r}$ at a temperature corresponding to $\beta$ -- see the black contours in figure \ref{fig:threads}(a).
An intuitive consideration of paths in 5D allows one to include some quantum correlations by considering paths beyond rings. It will be shown that these also reduce to ring configurations, and these many-body paths can lead to a second order transition as Feynman found for superfluid helium \cite{Feynman1953a, Feynman1953b}.

To see that many-body non-ring paths are equivalent to rings, consider two particles at positions ${\bf r}$ and ${\bf r}^\prime$, respectively, both at high temperature ($s = 0$). One can ask what the probability is of still finding two particles at positions ${\bf r}$ and ${\bf r}^\prime$ at a temperature corresponding to $\beta$. The various paths the two particles can take in order to contribute to the two-particle density can be seen in figure \ref{fig:threads}(b). This argument can be extended to any number of particles, but is here confined to the pair density for simplicity. Following figure \ref{fig:threads}(b), the unnormalized pair density will be
\begin{equation}
n({\bf r},{\bf r}^\prime,\beta) \propto q({\bf r},{\bf r},\beta) q({\bf r}^\prime,{\bf r}^\prime,\beta) +q({\bf r},{\bf r}^\prime,\beta) q({\bf r}^\prime,{\bf r},\beta)  .  \label{n2a}
\end{equation}
Integrating (\ref{n2a}) over all possible pairs of particles should give the total number of pairs
\begin{equation}
\int\int n({\bf r},{\bf r}^\prime,\beta) d{\bf r}d{\bf r}^\prime   = \frac{N(N-1)}{2}   \label{n2norm}
\end{equation}
which allows the proportionality constant for (\ref{n2a}) to be set to
\begin{equation}
k = \frac{N(N-1)}{2 Q^{(2)}(\beta)}  \label{kprop}
\end{equation}
where
\begin{eqnarray}
Q^{(2)}(\beta) &=&  \int\int \left[ q({\bf r},{\bf r},\beta) q({\bf r}^\prime,{\bf r}^\prime,\beta) +q({\bf r},{\bf r}^\prime,\beta) q({\bf r}^\prime,{\bf r},\beta) \right] d{\bf r}d{\bf r}^\prime  \nonumber \\
&=& \int  q({\bf r},{\bf r},\beta) d{\bf r} \int q({\bf r}^\prime,{\bf r}^\prime,\beta)d{\bf r}^\prime +\int d{\bf r}\left[ \int d{\bf r}^\prime q({\bf r},{\bf r}^\prime,\beta) q({\bf r}^\prime,{\bf r},\beta) \right] \nonumber \\
&=& Q(\beta)^2 + Q(2\beta) .  \label{Q2}
\end{eqnarray}
The last line of (\ref{Q2}) follows from (\ref{Q1}) and the fact that 
\begin{equation}
 \int d{\bf r}^\prime q({\bf r},{\bf r}^\prime,\beta) q({\bf r}^\prime,{\bf r},\beta) = q({\bf r},{\bf r},2\beta) .  \label{q2beta}
\end{equation}
Equation (\ref{q2beta}) shows that these cross-paths are equivalent to ring paths of twice the contour length, as first pointed out by Feynman \cite{Feynman1953b}. Equations (\ref{n2a})-(\ref{q2beta}) allows one to write the pair density as 
\begin{equation}
n({\bf r},{\bf r}^\prime,\beta) = \frac{N(N-1)}{2\left[Q(\beta)^2 + Q(2\beta)\right]} \left[q({\bf r},{\bf r},\beta) q({\bf r}^\prime,{\bf r}^\prime,\beta) +q({\bf r},{\bf r}^\prime,\beta) q({\bf r}^\prime,{\bf r},\beta)\right]  . \label{n2b}
\end{equation}
One gets the single-particle density by integrating over one of the position variables according to
\begin{equation}
n({\bf r},\beta) = \frac{2}{(N-1)}\int n({\bf r},{\bf r}^\prime,\beta) d{\bf r}^\prime .  \label{n1a}
\end{equation}
The pre-factor in (\ref{n1a}) is to switch from counting pairs to counting singlets. This gives
\begin{equation}
n({\bf r},\beta) = \frac{N}{\left[Q(\beta)^2 + Q(2\beta)\right]}\left[q({\bf r},{\bf r},\beta) Q(\beta) + q({\bf r},{\bf r},2\beta) \right] . \label{n1b}
\end{equation}
In the limit of low temperatures $\beta \rightarrow \infty$, numerical calculations show that the density saturates with respect to $\beta$ and so stops changing with $\beta$. In particular, it is found that 
\begin{eqnarray}
\lim_{\beta \rightarrow \infty}  Q(2\beta) &=& Q(\beta)^2   \label{Q2inf} \\
\lim_{\beta \rightarrow \infty} q({\bf r},{\bf r},2\beta) &=& q({\bf r},{\bf r},\beta) Q(\beta) . \label{q2inf}
\end{eqnarray}
Substituting (\ref{Q2inf}) and (\ref{q2inf}) into (\ref{n1b}) immediately gives the ring polymer density expression (\ref{dens1}). For some exotic systems, such as the superfluid helium studied by Feynman \cite{Feynman1953a, Feynman1953b}, there is a second order transition due to the correlations in (\ref{n1b}). Since the helium system is bosonic, one should also include three and higher-body terms in (\ref{n1b}). For fermions, due to the Pauli exclusion principle, the maximum number of particles in any state will be two, and so (\ref{n1b}) would not require any higher terms, and one would not expect to see a transition.\footnote{It may be possible to show in future work that the superconducting transition is related to quantum correlations in (\ref{n1b}), but for now this is speculation.} For example, even using a temperature on the order of the surface of the sun, one gets less than a one one-thousandth percent change in the SCFT calculated helium atomic binding energy due to non-ring (longer ring) paths. In order to intuitively picture phenomena using the 5D holographic principle, equation (\ref{diff1}) for open threads will be used instead of rings since this will give qualitatively equivalent results. Several archetypal quantum phenomena are now examined using the open thread picture. 

\subsection{Wave functions, quantum statistics and the measurement problem}
The non-ring paths shown in figure \ref{fig:threads}(a) by the grey contours and expressed in equation (\ref{n1b}) correspond exactly to matrix elements in 4D wave function QM arising from symmetric statistics \cite{Feynman1953b}. Anti-symmetric statistics requires replacing the plus sign in equation (\ref{n2a}) with a minus sign, as  should be done when replacing Feynman's discussion of bosons with anti-symmetric wave functions for fermions \cite{Feynman1953b}. In 5D, the minus sign in (\ref{n2a}) causes the mathematics to break down at zero temperature because the partition function corresponding to equation (\ref{Q2}) becomes equal to zero. This makes sense for two reasons: First, the propagator $q({\bf r},{\bf r},\beta)$ can be expanded in terms of complex orbitals that correspond to quantum mechanical states \cite{LeMaitre2023}. Since fermions cannot multiply occupy a single state (ignoring spin), the number of microstates available for such a configuration should be zero, as is found. In other words, from the 5D perspective, Pauli exclusion is the cause of anti-symmetric wave function statistics and not the effect. Second, the 5D holographic statement of QM does not use wave functions (or for that matter, quantum numbers or superpositions), and so concepts of bosons and fermions, which are based on wave function statistics, are superfluous. The theorems of DFT guarantee that all predictions of polymer SCFT applied to quantum particles must be the same as those of non-relativistic QM, and particles will turn out to behave in 4D as bosons or fermions based on their intrinsic and interaction properties. Threads in 5D need not be assigned a priori to either group to perform SCFT calculations. Likewise, since there are no wave functions, one can see from equation (\ref{dens1}) for the density, for example, that there is no wave function collapse during calculations. The measurement problem is an artifact of doing calculations in 4D for a system that can be rigorously viewed as projected out of 5D.

\subsection{Quantum kinetic energy and the uncertainty principle}
An extended object like a Gaussian thread has properties which arise from its internal degrees of freedom. Thermodynamically, these aspects are captured by the conformational entropy, which counts the number of configurations available to a thread when subject to a field $w({\bf r},\beta)$ at a temperature $T$. In SCFT, the contribution of the conformational entropy to the free energy of a system is given by \cite{Thompson2019}
\begin{equation}
F_{\rm conf} = -\frac{1}{\beta} \int d{\bf r} n({\bf r}) \left[\ln  q({\bf r},{\bf r},\beta)+\beta w({\bf r})\right] . \label{Fconf}
\end{equation}
The complete SCFT free energy expression for a quantum mechanical system is given in reference \cite{Thompson2019}\footnote{Equation (17) of reference \cite{Thompson2019} contains typos which are corrected here.} and accounts for all thermodynamic quantities except the quantum kinetic energy, while having the extra term (\ref{Fconf}) for conformational degrees of freedom of a classical thread in 5D. Thus, since the theorems of DFT guarantee that the SCFT expression must give the same results as QM, and exactly correct results for systems like the hydrogen atom and the Hartree-Fock helium atom are obtained, one is forced to identify the 5D conformational entropy as the 4D quantum kinetic energy.

The thread-like internal degrees of freedom are also the 5D expression of the 4D uncertainty principle \cite{Thompson2019, Thompson2020}. The derivation of the 5D governing diffusion equation (\ref{diff1}) in appendix A of reference \cite{Thompson2019} depends crucially on the uncertainty principle. In terms of position and momentum, if these quantities commute, then (\ref{diff1}) collapses to the thermodynamics of a point-like particle subject to the field $w({\bf r},\beta)$ rather than a random walk. In other words, equation (\ref{diff1}), which expresses the Gaussian thread nature of quantum particles in 5D, is equivalent to including the uncertainty principle in 4D. There is a close connection therefore between the uncertainty principle, quantum kinetic energy and conformational entropy which the holographic principle makes clear. 

\subsection{Stability of atoms}
The conformational entropy of SCFT threads also explains the stability of atoms. Classically, it is energetically favourable for electrons to be attracted into the Coulomb potential of the ionic core, collapsing the electron density into a spike at the nucleus. The conformational entropy, given by equation (\ref{Fconf}) however, would become enormous because, even at zero temperature, the thread would be confined to a single microstate. SCFT reveals that the experimentally observed electron density is a result of the frustration between energy and conformational entropy in the 5D statistical mechanical picture.

\subsection{Tunnelling}
Consider the operator on the right hand side of the SCFT diffusion equation (\ref{diff1}), which is
\begin{equation}
\mathcal{H} \equiv \frac{\hbar^2}{2m} \nabla^2 - w({\bf r},\beta)	.   \label{H}
\end{equation}
This obeys an eigenvalue equation
\begin{equation}
\mathcal{H} \phi_i({\bf r}) = E_i \phi_i({\bf r})       \label{eig1}	
\end{equation}
where $\phi_i({\bf r})$ and $E_i$ are the eigenfunctions and eigenvalues, respectively. Note that whereas $q({\bf r}_0,{\bf r},s)$ was completely real and positive definite, $\phi_i({\bf r})$ can be complex. Equation (\ref{eig1}) is a one-particle time-independent Schr\"{o}dinger equation, and is called the Kohn-Sham equation in the context of DFT. The density expression (\ref{dens1}) can likewise be shown to become identical to the Kohn-Sham density expression, assuming perfect enforcement of the Pauli exclusion principle -- see appendix B of reference \cite{Thompson2019} and appendix C of reference \cite{LeMaitre2023}. For the case of a single quantum particle, the SCFT -- Kohn-Sham DFT duality holds without conditions. Thus, the SCFT equations formed around (\ref{diff1}) will make all the same predictions as static quantum mechanics, including tunnelling. This is not surprising since ring polymer formalisms are used in quantum simulations to study tunnelling \cite{Althorpe2009}.

\begin{figure}
\includegraphics[width=1.0\textwidth]{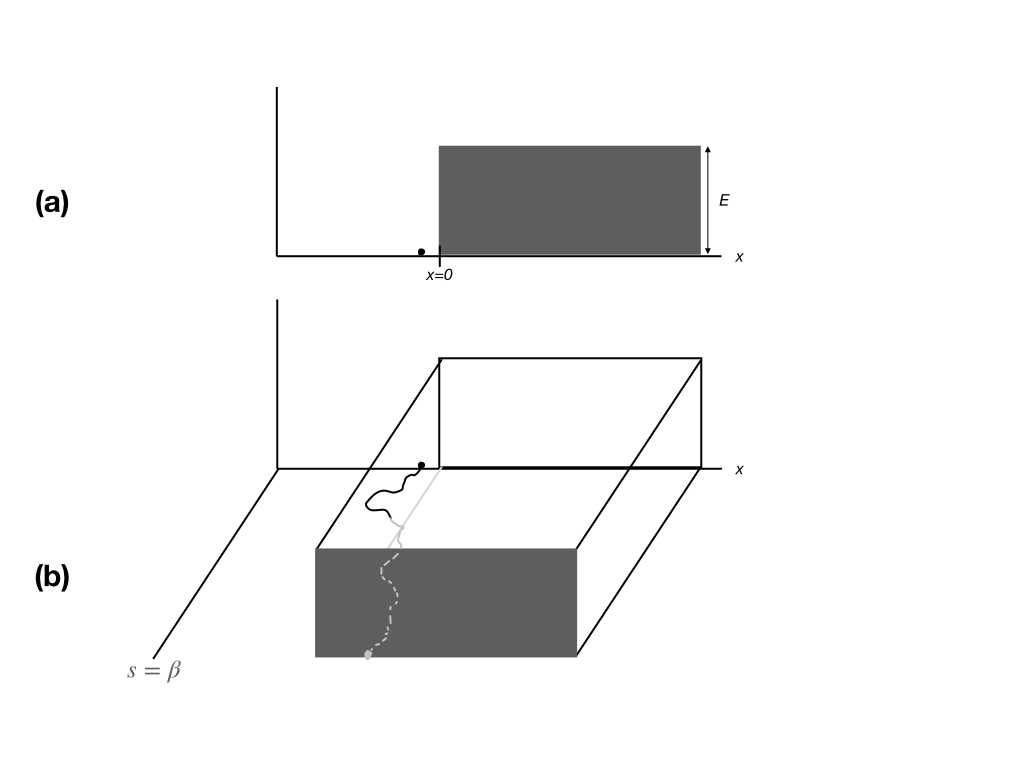}
\caption{Schematic of static tunnelling in the holographic picture. (a) A classical particle, shown by a dot, cannot enter the energy region $E$ if it has too low a kinetic energy. (b) A thread in thermal-space starts at the same position as the classical particle, where it also cannot enter the region $E$. As the contour extends from the initial point according to (\ref{diff1}), the barrier $E$ discourages, but does not forbid entry.} \label{fig:tunnelling}
\end{figure}

Equation (\ref{diff1}) can therefore be used to qualitatively understand tunnelling in the holographic 5D context. Figure \ref{fig:tunnelling}(a) shows a typical tunnelling situation with a barrier of finite energy $E$. 
A classical particle with insufficient kinetic energy cannot penetrate the barrier. For a 5D thread described by (\ref{diff1}), $w({\bf r},\beta) = E$ in the range $x>0$ represents an energy penalty that discourages the contour from entering that region, but does not forbid it -- there is a competing entropy benefit which lowers the free energy due to increased conformations when the thread enters the region -- see figure \ref{fig:tunnelling}(b). At $s=\beta$, there is thus a non-zero probability of finding the particle in the barrier. As long as the initial position of the thread at $s=0$ is outside the barrier, the extra thermal dimension provides a classical trajectory through which to enter the barrier zone.  

\section{Dynamic Properties}

It is not immediately obvious how to write down the mathematical formulation for the dynamics of quantum particles in 5D as classical thermal threads. This is not surprising, since writing a dynamical version of SCFT for actual polymers is similarly difficult and one is forced to use approximations. However, just as it is known that chemical polymers continue to be extended one dimensional objects even though one cannot write exact SCFT mathematical equations for their dynamics, so it can be shown that quantum particles can continue to be represented as 1D thermal-world-lines even without giving explicit dynamic SCFT equations.

To see this, the argument of reference \cite{Thompson2022} is followed, considering again the eigenvalue equation of the operator on the right hand side of (\ref{diff1}). As mentioned, the Kohn-Sham equation (\ref{eig1}) is a one-particle time-independent Schr\"{o}dinger equation.  In 4D wave function QM, non-relativistic dynamics are commonly included through postulating the time-dependent Schr\"{o}dinger equation. It is therefore convenient to do a similar thing with the Kohn-Sham DFT expression (\ref{eig1}), that is, it is generalized and dynamics are postulated to be given by the equation 
\begin{equation}
i\hbar \frac{\partial}{\partial t} \phi_i({\bf r},t) = \mathcal{H} \phi_i({\bf r},t). \label{SE}
\end{equation}
Together with the corresponding density and field expression -- see reference \cite{Thompson2022} -- equation (\ref{SE}) is identical to the formulas of time-dependent DFT (TDDFT). A result of TDDFT is that the time-dependent density $n({\bf r},t)$ is proportional to a real quantity which is the time-dependent version of $q({\bf r},{\bf r},\beta)$ \cite{vanLeeuwen2006b}:
\begin{equation}
n({\bf r},t) \propto q({\bf r},{\bf r},\beta,t) .\label{nqt}
\end{equation}
Just as $q({\bf r},{\bf r},\beta)$ is the mathematical expression of thermal-world-line statistics in the static case, (\ref{nqt}) demonstrates that the classical thread picture of quantum particles in 5D is consistent with quantum dynamics, even if on a practical level, one uses the equations of TDDFT, including (\ref{SE}), as a black-box to perform calculations in terms of complex eigenfunctions. 

The consistency of the 5D polymer thread model with TDDFT means that, through the Runge-Gross theorem \cite{Runge1984}, all dynamic predictions of the 5D thread picture must be consistent with those of the dynamic 4D wave function picture. Similar to the static case, one expects to be able to use ring polymers to effectively describe quantum particles, but for explanatory purposes, open threads based on equation (\ref{diff1}) will be used instead. The holographic principle allows one to consider a number of dynamic effects in non-relativistic QM.

\subsection{The double slit experiment}
\begin{figure}
\includegraphics[width=1.0\textwidth]{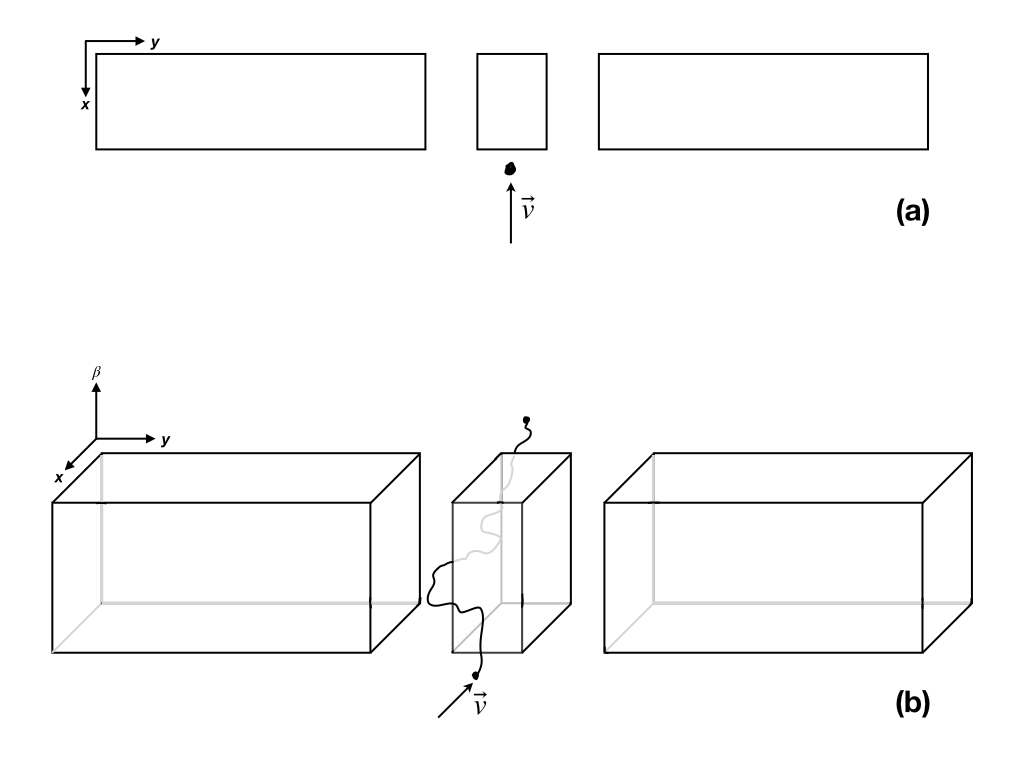}
\caption{A schematic of the double slit experiment. (a) A classical particle, denoted as a dot and with a velocity ${\bf v}$, is shown incident on the double slit  at a position such that it cannot pass through either slit; it will therefore be blocked from being detected on the final screen (not shown). (b) A thread in thermal-space with $s=0$ at the same position as the classical particle in (a). The probability of the particle being found at this position for $s=\beta$ is not 1; there is a non-zero probability that the other end of the contour will be found at a position that would be impossible for a classical point particle.}
 \label{fig:doubleslit}
\end{figure}

A quasi-static approximation can be adopted in order to intuitively understand interference in the double slit experiment. The static governing equation (\ref{diff1}) requires that if a classical particle is at a position ${\bf r}$ at high temperature, then in the absence of other particles or external influences, the probability of finding the particle at a temperature where quantum effects are important will be a Gaussian distribution centred on ${\bf r}$. Microscopically, this means that each particle thread will explore various thermal paths and might be found at a position that would be impossible classically by wrapping around infinite potential obstructions in the thermal dimension-- see figure \ref{fig:doubleslit}.
As discussed in references \cite{Thompson2019} and \cite{Thompson2022}, the static non-locality of threads means that the density of particles predicted at the screen will be different from that expected classically, but without a further dynamic aspect, it does not yet agree with experimental predictions. It is known from TDDFT and the Runge-Gross theorem that 5D classical threads must be able to reproduce interference phenomena -- this is consistent with equation (\ref{SE}) which has an oscillatory nature. In other words, the propagator $q({\bf r},{\bf r},\beta,t)$ for the probability of a thread returning to its starting point must change in time commensurate with the de Broglie wavelength, or the function $q({\bf r}_0,{\bf r},\beta,t)$ for the probability of a thread ending at a different position, for a fixed ${\bf r}_0$ and $t$, will show the de Broglie wavelength. Much like an oscillating pendulum has a non-uniform probability of being found at various points along its period, so a pulsating 5D thread at an instant of time should oscillate with its high temperature initial point ${\bf r}_0$ at $s=0$ fixed and other points $s$ free to vibrate according to (\ref{SE}) -- see figure \ref{fig:threads}(a). By construction then, a classical 5D thread must agree with experimental double slit results since the time-dependent Kohn-Sham equation (\ref{SE}) is identical with the time-dependent Schr\"{o}dinger equation for a single particle, while still being consistent with classical 5D thread statistics through (\ref{nqt}).

\subsection{Geometric phase and the Aharonov-Bohm effect}
When a quantum particle is transported through a closed loop in a parameter space, a phase change results due to the non-flat geometry of the space \cite{Berry1984}. This phase, known as the geometric phase or Berry's phase, is detectable in experiments, a famous example being the Aharonov-Bohm effect \cite{Ehrenberg1949, Aharonov1959}, but is not unique to quantum systems. It is seen in classical phenomena such as parallel transport and the precession of classical pendulums, which also experience phase changes when transported through closed loops in non-flat geometries. The 5D holographic principle in which quantum particles are classical threads oscillating in thermal-space-time, as described in the previous section, immediately allows a 5D classical interpretation of quantum geometric phase phenomena, such as the Aharonov-Bohm effect, in terms of classical geometric phase changes in systems like pendulums. 

\subsection{Entanglement}
\begin{figure}
\includegraphics[width=1.0\textwidth]{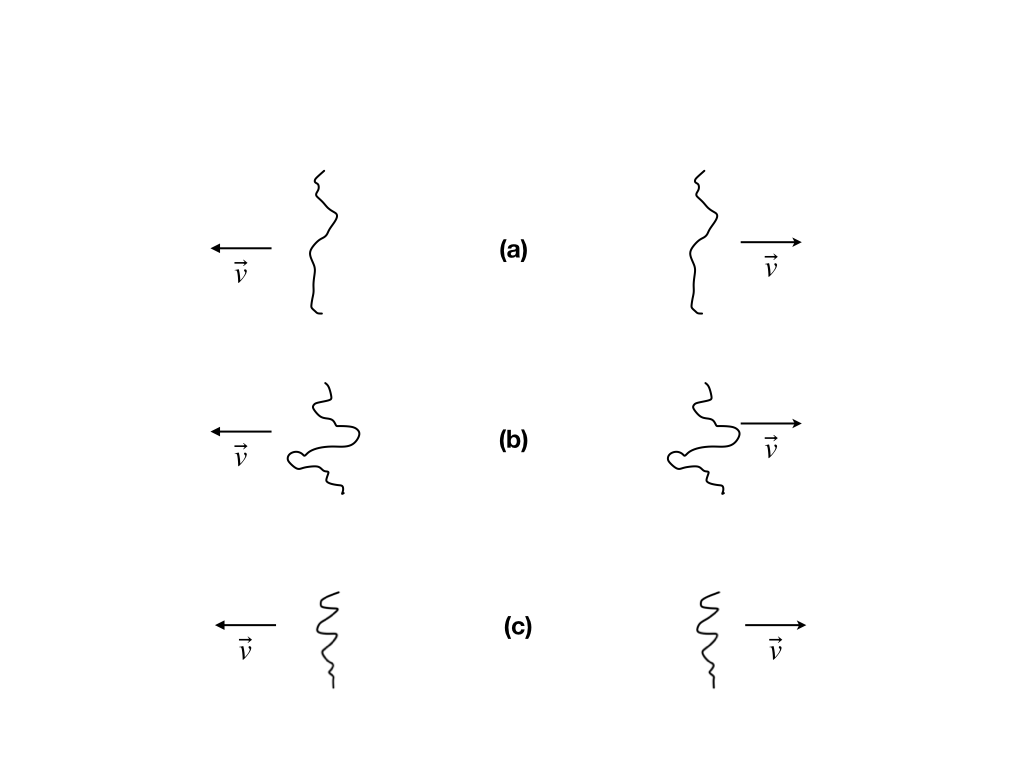}
\caption{Three examples, (a)-(c), of 5D threads moving in opposite directions with the same conformations. The conformations change as the particles travel, but each thread in a pair will maintain identical conformations at any given time.}
 \label{fig:entangled}
\end{figure}

Imagine a quantum particle which is split into two daughter particles which travel off in opposite directions. The 5D holographic principle view of this is shown in figures \ref{fig:entangled} (a)-(c).
In each case, the two daughter threads can be considered to be entangled because their internal degrees of freedom (the conformational shapes they have) are correlated (they have the same shape as each other). With no further communication between them, any experiments done on one will be correlated with the other, to a larger extent than would be expected for classical point particles. This can be quantitatively seen with SCFT if measurements of positions and momenta for pairs of entangled particles at distant locations are considered. 

The Fourier transform of the governing thread equation (\ref{diff1}) gives a relationship between momentum and position -- see appendix A of reference \cite{Thompson2019}. Suppose Alice and Bob each perform a series of momentum or position measurements on opposite sides of the experimental apparatus. Given that the threads emitted in the two directions have identical conformations,\footnote{The conformations would change as the particles travel, but they would change in exactly the same way, so they would always have the same conformation as each other at any given time.} if Bob on his side of the apparatus happens to have chosen to measure the same quantity as Alice, they will completely agree on the results they find. However, if Alice measures momentum and Bob measures position, after they have collected enough data, it will seem as if the wave function of the entire system collapses whenever one of them takes a measurement only on their own side, even though the 5D picture does not use wave functions.

To see this, suppose Alice performs a series of momentum measurements on her side of the experimental setup, and she makes note of all momenta that fall within a narrow range -- that is she notes all particles that have a chosen momentum. If Bob happens to have measured position for those pairs, when they compare their results, Alice will have one value of momentum for all her particles, but Bob will have essentially random values of position. That is because although each particle in a pair has the same thread conformation as its partner, separate pairs will have completely different conformations between them -- see figure \ref{fig:entangled}. Since the polymeric threads are not point particles, the classical relationship between momentum and position is not valid. From the Fourier transform of equation (\ref{diff1}), in the absence of a field $w({\bf r},\beta)$, a known single momentum will give a distribution of positions -- the actual position found for a single experiment will depend on the specific conformation of the polymer. The quantitative values of the correlations for 5D classical thread pairs must agree with those of 4D quantum mechanics from the theorems of DFT.

This guess for the mechanism of entanglement in 5D is very speculative and includes only the maximally and minimally correlated situations, which Bell comments are ``... the only features ... commonly used in verbal discussions of this problem'' \cite{Bell1964}. He notes that these features are easily explained in terms of local hidden variables. In order to completely satisfy Bell-type theorems with hidden variables, intuitively unacceptable non-local variables are needed. What is unacceptable in 4D may be more palatable in higher dimensions, and an extended mechanism involving cross-paths, equivalent to a single double-length contour that must pass through both Alice and Bob's measuring devices, offers an intriguing mechanism not restricted by the speed of light since the imaginary time plane is not Lorentzian. 

\subsection{Other features}
The holographic principle for QM works in a 5D thermal-space-time. It has also been shown that it obeys a cylinder condition; specifically, in the classical limit, the thermal dimension still exists, but no longer has any quantum effect on predictions -- see appendix C of reference \cite{Thompson2019}. Five dimensions subject to a cylinder condition in the classical limit are the assumptions of Kaluza theory \cite{Kaluza1921, Wesson1997}, and it follows that the postulates necessary for the polymer thread picture of QM include those necessary to derive electromagnetism from the structure of general relativity by using five dimensions instead of four \cite{Thompson2022}. Thus, the 5D holographic principle of QM, in addition to having fewer postulates than 4D wave function theory, and avoiding various QM pathologies, leads directly to other physical phenomena outside of QM without additional assumptions \cite{Thompson2022}.

There are many other quantum phenomena that one could attempt to intuitively explain, that have not yet been addressed. For example, the physical origins of spin in 5D, interaction-free measurement (the Elitzur-Vaidman bomb experiment) \cite{Vaidman1993}, other interferometric predictions of QM \cite{Vaidman2020, Vaidman2022}, non-statistical tests of entanglement, such as Greenberger–Horne–Zeilinger states, etc. While the holographic principle will not always be as easy to apply as in the examples listed in this paper, the theorems of DFT guarantee that the 5D classical thread model will make exactly the same predictions as standard non-relativistic QM.

\section{The Pauli Exclusion Principle}
As has been mentioned, SCFT can be shown to be equivalent to quantum DFT assuming enforcement of the Pauli exclusion principle \cite{Thompson2019}. This is what allows the use of the theorems of DFT to show the holographic principle connecting 5D polymer SCFT with 4D wave function QM. It is therefore necessary to postulate the nature of the exclusion principle in the classical 5D picture, in addition to the two other assumptions, namely that quantum particles are classical threads in 5D, and that these threads vibrate according to (\ref{SE}) \cite{Thompson2022}. 

Quantum threads in 5D are postulated to obey excluded volume -- just as the trajectories of classical particles do not allow multiple particles to be in the same place at the same time, so it is assumed that threads cannot be in the same place at the same imaginary time (same value of $\beta$). This property for classical threads in 5D maps onto the Pauli exclusion principle for particles in 4D. 

There are several reasons for assuming this excluded volume: First, it is already an accepted feature of the quantum-classical isomorphism that gives correct results. Feynman used the excluded volume of threads in imaginary time to justify removing trajectories in his study of the $\lambda$-transition \cite{Feynman1953b}. He did not identify this with the exclusion principle -- he was working with bosons -- but following his example, any massive particle would have to have this excluded volume feature, including electrons. The mystery in terms of the electron is why do up to two of these fermions, assuming opposite spins, \emph{not} feel excluded volume in the 5D space? This question has not been answered, but this feature can be accepted as a property of electrons since, although quantum particles are given structure as one-dimensional threads, this does not say anything about the cross-sectional structure of each thread -- just like in 4D QM, electrons have no internal structure. Using 5D excluded volume in this way for electron calculations produces correct shell structure of the atoms \cite{Thompson2020, LeMaitre2022} and correct molecular bonding \cite{Sillaste2022}, at least within the context of the approximations used to implement the excluded volume.

A second reason that supports 5D classical thread excluded volume is that it gives the correct scaling behaviour in the uniform limit for both the quantum kinetic energy and Dirac exchange. Thomas \cite{Thomas1927} and Fermi \cite{Fermi1927} found that the quantum kinetic energy of the high density uniform electron gas should scale with the density of the gas $n$ as $n^{5/3}$. Dirac \cite{Dirac1930} found a correction term due to exchange effects that scales as $n^{4/3}$. If techniques of polymer scaling theory as given by de Gennes \cite{deGennes1979} are used, assuming excluded volume between threads in 5D gives both the correct Thomas-Fermi expression \cite{Thompson2020, LeMaitre2022} and the correct Dirac exchange term \cite{LeMaitre2022}.

Third, the 5D excluded volume picture approaches satisfying conditions necessary for the Pauli field as described by Levy and Ou-Yang \cite{Levy_Ou-Yang1988.article}. To date, the excluded volume model has only been implemented on a coarse approximate level \cite{LeMaitre2022, LeMaitre2023}, but even this approximate version satisfies four of five conditions, with the fifth being violated by an amount commensurate with the scale of the approximation -- more details can be found in references \cite{LeMaitre2022} and \cite{LeMaitre2023}. More work is needed to further test the postulate relating excluded volume in 5D to the Pauli exclusion principle in 4D.

\section{Conclusions and Future Work}  \label{Conclusions}

A classical thread model for quantum particles in 5D can produce all the same predictions as 4D quantum mechanics, but with fewer postulates. These axioms are: 1. Quantum particles are classical threads in 5D (which should be formally treated as rings, although that condition has been relaxed in this work for illustrative purposes). 2. Time evolution is governed by the time-dependent Kohn-Sham equation, which can be shown to be consistent with the statistics of 5D classical threads. 3. Pauli exclusion is enforced through higher dimensional excluded volume. 

The duality between 5D classical threads and 4D quantum wave functions defines a holographic principle. Advantages of the 5D perspective, in addition to having fewer postulates, include: a realist model which is, in the context of higher dimensions, deterministic;  no measurement problem; an explanation of randomness through the ensemble interpretation utilizing the internal conformational degrees of freedom of 5D threads within classical statistical mechanics; an interpretation of quantum statistics, exchange and the Pauli exclusion principle; intuitive explanations for quantum phenomena such as the Aharonov-Bohm effect,  the uncertainty principle, quantum kinetic energy, tunnelling and the double slit experiment.

Many issues remain unexplored however. Spin is easily included in 5D SCFT calculations, but no physically intuitive picture for it has been provided yet -- for example, no explanation is given as to why two electrons of opposite spins do not feel excluded volume in 5D. Non-statistical tests of entanglement, such as Greenberger–Horne–Zeilinger states, should be considered. A relativistic version of the principle is required, and connections with thermal quantum field theory, related to Matsubara imaginary time, should be explored \cite{Vitiello2011}. Justifications that the Pauli exclusion principle is excluded volume in 5D have been listed, but further evidence is necessary. Improved approximations of the 5D excluded volume in SCFT calculations could show whether the shell structure of atoms approaches chemical accuracy, or diverges from experiment. Likewise, more complex SCFT molecular calculations and time-dependent SCFT problems could either support or contradict the 5D picture. One advantage of 5D SCFT is that it is readily compared to experiments, and although inevitable approximations in numerical calculations are a confounding factor, this limitation is no different from any other quantum calculation. If nothing else, the SCFT holographic principle is a useful quantum calculational tool, but its implications for quantum foundations are much more interesting.

\section*{Acknowledgements}

The author acknowledges helpful discussions with many of the participants of the 15th Biennial Quantum Structure Conference, with P. A. LeMaitre for suggesting changes to the manuscript, and with M. W. Matsen who pointed out the relationship (\ref{q2beta}) between cross-path contours and longer single rings. 

\bibliography{DFTbibliography}

\end{document}